\documentclass[12pt]{article}
\usepackage{graphicx}
\usepackage{epsfig}
\begin {document}
\begin{center}
\bf ELASTIC pp SCATTERING AT LHC ENERGIES

\vspace{.2cm}

C. Merino$^*$ and Yu.M. Shabelski$^{**}$ \\

\vspace{.5cm}
$^*$ Departamento de F\'\i sica de Part\'\i culas, Facultade de F\'\i sica 
\\ 
and Instituto Galego de F\'\i sica de Altas Enerx\'\i as \\ 
Universidade de Santiago de Compostela \\
Santiago de Compostela 15782 \\
Galiza, Spain \\
E-mail: merino@fpaxp1.usc.es \\

\vspace{.2cm}

$^{**}$ Petersburg Nuclear Physics Institute \\
Gatchina, St.Petersburg 188350, Russia \\
E-mail: shabelsk@thd.pnpi.spb.ru
\vskip 0.9 truecm

\vspace{1.2cm}

\end{center}

We consider the first LHC data for elastic $pp$ scattering in the framework 
of Regge theory with multiple Pomeron exchanges. The simplest eikonal 
approach allows one to describe differential elastic cross sections 
at LHC, as well as $pp$ and $\bar{p}p$ scattering at lower collider energies, 
on a reasonable level.  

\vskip 1.5cm


\newpage

\section{Introduction}

In Regge theory the Pomeron exchange dominates the high energy soft
hadron interaction. The Pomeron has vacuum quantum numbers, so the 
difference in $pp$ and $\bar{p}p$ should disappear. At LHC energies
the contributions of all other exchanges to the elastic scattering amplitude
becomes negligible, and then one can directly extract the Pomeron parameters 
from the experimental data.

In the present paper we consider the first LHC data (TOTEM 
Collaboration~\cite{TOT1a}) for $pp$ small angle elastic scattering and we 
compare them with the simplest approaches of Regge theory and with the results
 ed for other lower collider energies.

The experimental elastic cross section is well described by a pure 
exponential form in the interval of momentum transfer 
$\vert t \vert = 0 - 0.3$ GeV$^2$. In this interval the cross section falls 
down more than 400 times. The  experimental ratio of $\sigma_{el}/\sigma_{tot}$
is equal to $\sim 0.25$ and an intersting point to be analysed is whether in 
the framework of a conventional Regge theory we have a chance to describe 
such a large elastic cross section without introducing, either a second 
Pomeron pole with a large intercept $\alpha_P(0)=1.362$, as in~\cite{DL}, a 
rather non-trivial spatial $b_t$-distribution of the matter in the proton 
with a deep minimum at $b_t=0$, like it was done in~\cite{UG} (see the form 
of $\gamma(b)$ in Eq.~(9) of~\cite{UG})\footnote{Note that the total cross 
section obtained in~\cite{UG} for $\sqrt s=7$~TeV is $\sigma_{tot}=90.9$~mb, 
much smaller than that measured by TOTEM.}, or more complicated approaches,
such as the three-channel eikonal model~\cite{RMK} or the model~\cite{Sel} in 
which
uses the general parton distributions.

\section{Elastic Scattering Amplitude at LHC energies}

Let us consider elastic $pp$ ($\bar{p}p$) scattering at very high energies
in the framework of Regge-Gribov theory~\cite{Gri}, where only Pomeron 
exchanges should be accounted for. It is suitable to use the following 
normalization of the elastic scattering amplitude $A(s,t)$:
\begin{equation}
\sigma^{tot} = 8 \pi\cdot Im A(s,t=0)\;, \;\;
\frac{d\sigma}{dt} = 4\pi\cdot \vert A(s,t) \vert^2 \;.
\end{equation}

The simplest contribution to the elastic scattering amplitude is the 
one-Pomeron, $P$, exchange, that can be written as:
\begin{equation}
A^{(1)}(s,t) = \gamma(t) \cdot \left(\frac{s}{s_0}\right)^{\alpha_P(t) - 1}
\cdot \eta(\Theta) \;,
\end{equation}
where $\gamma(t) = g_1(t)\cdot g_2(t)$, $g_1(t)$ and $g_2(t)$ are the 
couplings of a Pomeron to the beam and target hadrons, 
$\alpha_P(t) = \alpha_P(0) + \alpha'_P\cdot t$ is the Pomeron trajectory, 
$\alpha_P(0)$ (intercept) and $\alpha'_P$ (slope) are some numbers, and 
$\eta(\Theta)$ is the signature factor which determines the complex 
structure of the scattering amplitude ($\Theta$ equal to +1 and to $-1$ for 
Reggeon with positive and negative signature, respectively). Specifically for
Pomeron exchange ($\Theta = +1$):
\begin{equation}
\eta(\Theta) = \frac{1 + \Theta \cdot exp [-i \pi \alpha_P(t)]}
{\sin{[\pi \alpha_P(t)]}} = i - \tan^{-1} 
\left(\frac{\pi \alpha_P(t)}2\right) \;.
\end{equation}

In the case of a Pomeron trajectory with $\alpha_P(0) > 1$, the correct
asymptotic behavior $\sigma_{tot} \sim \ln^2s$ \cite{Volk,Kop} compatible
with the Froissart bound \cite{Froi}, can only be obtained by taking into
account the multipomeron cuts.

Indeed, for the Pomeron trajectory
\begin{equation}
\alpha_P(t) = 1 + \Delta + \alpha'_P\cdot t\;,\,\, \Delta > 0 \;,
\end{equation}
the one-Pomeron contribution to $\sigma^{tot}_{hN}$  rises with energy as
$s^{\Delta}$. To comply with the $s$-channel unitarity and, in particular, with
the Froissart bound, this contribution should be screened by the multipomeron
discontinuities shown in Fig.~1.
\begin{figure}[htb]
\centering
\vskip -3.8cm
\includegraphics[width=.6\hsize]{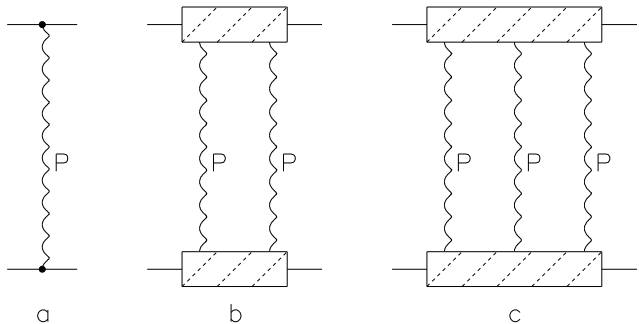}
\vskip -.6cm
\caption{\footnotesize
Regge-pole theory diagrams: (a) single, (b) double, and (c) triple
Pomeron exchange in elastic $pp$ scattering.}
\end{figure}

A simple quasi-eikonal treatment \cite{Kar1} allows one to present the
total elastic scattering amplitude $A(s,t)$ as a series 
\begin{equation}
A(s,t) = A^{(1)}(s,t) + A^{(2)}(s,t) + A^{(3)}(s,t) + ... \;,
\end{equation}
where each $A^{(n)}(s,t)$ contribution corresponds to the exchange of $n$ 
Pomerons. The value of $A^{(1)}(s,t)$ is given by Eq.~(2), and
\begin{equation}
A^{(2)}(s,t) = \frac1{2!} \int \frac{d^2\vec{q}_1}{\pi}\cdot 
A^{(1)}(s,\vec{q}_1) \cdot i \cdot A^{(1)}(s,\vec{q}-\vec{q}_1) \,
\end{equation}
\begin{equation}
A^{(3)}(s,t) = \frac1{3!} \int \frac{d^2\vec{q}_1}{\pi}\cdot 
\frac{d^2\vec{q}_2}{\pi}\cdot A^{(1)}(s,\vec{q}_1) \cdot i \cdot 
A^{(1)}(s,\vec{q}_2) \cdot i \cdot 
A^{(1)}(s,\vec{q}-\vec{q}_1-\vec{q}_2) \;,
\end{equation}
where all $q_i$ are two-dimensional vectors in the perpendicular plane to the
beam axis, $t = -{\vec{q}}^2$.  

The results of the integrations in Eqs.~(6), (7), etc., depend on the 
assumption about the form of the function $\gamma(t)$, with $t=-q^2$. These
integrations can be analytically performed in the simplest case of Gaussian
functions:  
\begin{equation}
\gamma(q^2) = \gamma_0\cdot e^{-R^2 \cdot q^2} \;.
\end{equation}
In this case the total elastic scattering amplitude of Eq.~(5) is equal to
\begin{equation}
A(s,t) = \eta_P\cdot\gamma_0\cdot e^{\Delta \xi}\cdot\sum_{n=1}^{\infty} 
\frac1{n\cdot n!} \left(i\cdot C\cdot\frac{\eta_P\cdot(q^2/n^2)\cdot\gamma_0}{\lambda}\cdot 
e^{\Delta\cdot\xi} \right)^{n-1} \cdot exp\left[-\frac{\lambda}n q^2\right] \;, 
\end{equation}
where $C$ is the quasi-eikonal enhancement coefficient (see~\cite{Kar1}), 
$\lambda = R^2 + \alpha'_P\cdot \xi$, $\xi = \ln{s/s_0}$, $s_0$ = 1~GeV$^2$.

At asymptotically high energies, $s \to \infty$, the amplitude of Eq.~(9)
leads to the Froissart behaviour of the total cross section,
$\sigma^{tot}(s) \sim \ln^2{s}$. 

On the other hand, it is well-known that the form of the function $\gamma(q^2)$
in Eq.~(8) is in contradiction with the experimental data on the shape of the differential
elastic cross section, so we have also used the parametrization of $\gamma(q^2)$ as a
sum of two gaussians:
\begin{equation}
\gamma(q^2) = \gamma_0\cdot(a \cdot e^{-R^2_1 \cdot q^2} + 
(1-a) \cdot e^{-R^2_2 \cdot q^2}) \;,
\end{equation}
that leads to a better agreement with the data. 

\section{Comparison with the experimental data}

The results of the calculation of $d\sigma/dt$ at $\sqrt{s}$ = 7~TeV, obtained
with the one-exponential para\-met\-rization of $\gamma(q^2)$ in Eq.~(8), are
presented in Fig.~2. The values of $\gamma_0$ were fixed by the value of 
$\sigma^{tot}$ at the same energy measured by TOTEM Collaboration 
\cite{TOT1a}. The two theoretical curves correspond to the values $C = 1.5$
(quasi-eikonal approach) and $C = 1$ (eikonal approach), and both are in total
disagreement with the experimental data (several experimental points presented
in Fig.~2 are taken from~\cite{TOT2}).

\begin{figure}[htb]
\centering
\includegraphics[width=.65\hsize]{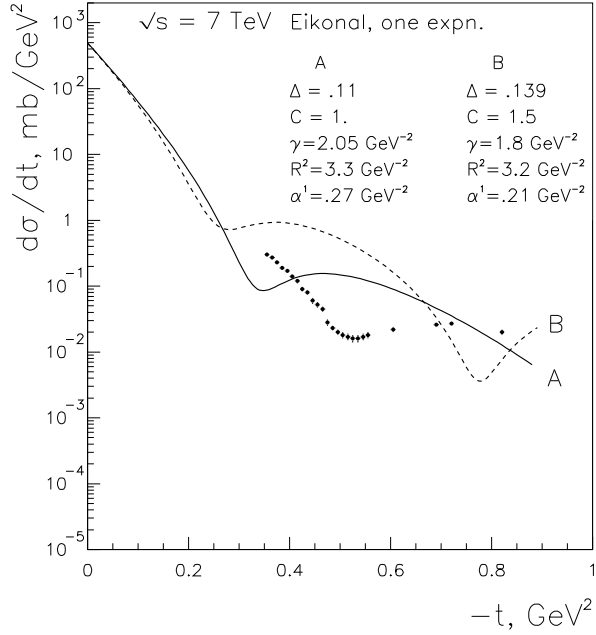}
\vskip -.6cm
\caption{\footnotesize
The differential cross section of elastic $pp$ scattering at $\sqrt{s}$ = 7 
TeV calculated in both eikonal (A) and quasi-eikonal (B) approaches, with the 
one-exponential parametrization of $\gamma(q^2)$ in Eq.~(8). The experimental 
points are taken from~\cite{TOT2}.}
\end{figure}

The main reason of the disagreements of the two theoretical curves in Fig.~2 
with the experimental data comes from the rather large rescattering 
contributions (exchanges of several Pomerons) in Eq.~(9). These contributions 
transform the bare Gaussian $t$-dependence of $d\sigma/dt$ given by Eq.~(8) 
into functions faster decreasing with $q^2$, whereas the experimental LHC 
data~\cite{TOT1a,TOT2} practically show a Gaussian $t$-dependence.

The simplest way to avoid this problem is to use a two-exponential form
for the function $\gamma(q^2)$ as the one given by Eq.~(10). All the integrals 
in Eqs.~(6), (7), etc., can be analytically calculated, giving an expression 
for $A^{(n)}$:
\begin{eqnarray}
A^{(n)}(s,q^2) & = & \frac{i^{(n-1)}}{n!}\cdot\left[\eta_P\cdot(q^2/n^2)
\cdot\gamma_0 e^{\Delta\cdot\xi} \right]^n\cdot\left [\frac{a^n}{n
\cdot\lambda_1^{(n-1)}} + \frac{(a-1)^n} {n\cdot\lambda_2^{(n-1)}} \right. + 
\\ \nonumber
& + & \sum_{k=1}^{n-1} C^k_n\cdot\frac{a^{(n-k)}\cdot(1-a)^k} 
{\lambda_1^{(n-k-1)}\cdot\lambda_2^{(k-1)}\cdot [k\cdot\lambda_1 + 
(n-k)\cdot \lambda_2]} \cdot \\ \nonumber
& \cdot & \left. exp\left(-\frac{q^2}{(n-k)\cdot\beta_1 + k\cdot\beta_2}
\right) \right] \;, \\ \nonumber
C^k_n & = & \frac{n!}{k!\cdot(n-k)!} \;, \;\; \lambda_i = R^2_i + 
\alpha'_P\cdot \xi \;, \;\; \beta_i = 1/\lambda_i \;.
\end{eqnarray}
  
\begin{figure}[htb]
\centering
\includegraphics[width=.65   \hsize]{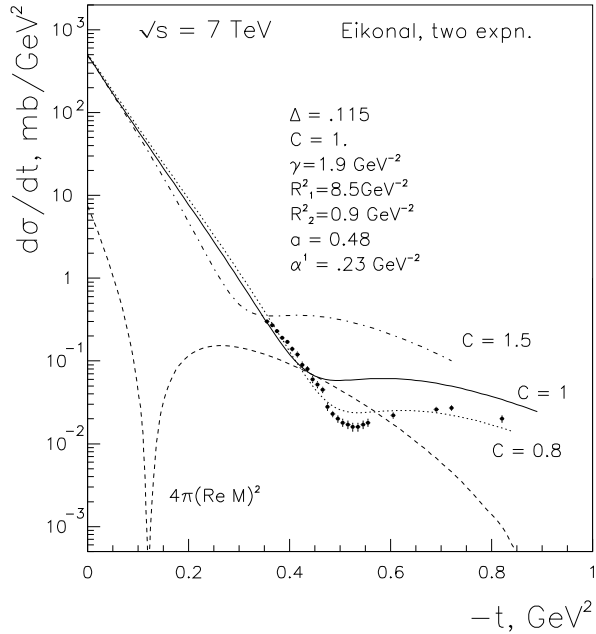}
\vskip -.6cm
\caption{\footnotesize
The differential cross section of elastic $pp$ scattering at $\sqrt{s}$ = 7 TeV
calculated in the eikonal approach, C = 1 (solid curve), and the contribution to
differential cross section of elastic $pp$ scattering at $\sqrt{s}$ = 7 TeV of
only the real part of the amplitude (dashed curve) by using  the two-exponential
parametrization of the function $\gamma(q^2)$ given in Eq.~(10). The results
 obtained for the differential cross section in the two quasi-eikonal 
approaches with C = 1.5 and C = 0.8 are also shown. The experimental points 
have been taken from~\cite{TOT2}.}
\end{figure}.

The results of the calculation of $d\sigma/dt$ at $\sqrt{s}$ = 7 TeV obtained 
with the parametrization of the function $\gamma(q^2)$ given in Eq.~(10) are
presented in Fig.~3. The quasi-eikonal case in which C = 1.5 leads again to
a too fast decrease and it gives a too small slope at low $q^2$. Instead, the
eikonal approach, C = 1, leads to a reasonable description of the data.
The agreement of our calculations with the experimental data~\cite{TOT1a} at
small $q^2$ comes from the facts that both the calculated and the experimental
$q^2$-dependences are close to Gaussians and that the calculated 
value of $\sigma^{tot}$ is in agreement with the experimental 
result~\cite{TOT1a} (see below). One important point to be stressed is that 
in the diffraction minimum, or in the beginning of the ``shoulder'', the 
cross section $d\sigma/dt$ is practically determined by only the real part of 
the amplitude (see solid and dashed curves in Fig.~3).

The solid curve in Fig.~3 was calculated with the following values of the 
parameters:
\begin{eqnarray}
\Delta & = & 0.115 , \;\;  \alpha'_P = 0.23 \; {\rm GeV}^{-2}, \;\;
\gamma = 1.9 \; {\rm GeV}^{-2}, \\ \nonumber 
a & = & 0.48, \;\; R^2_1 = 8.5 \; {\rm GeV}^{-2}, \;\; R^2_2 = 0.9 \;
{\rm GeV}^{-2} \;.
\end{eqnarray}

The quality of the description is even better in the quasi-eikonal case with 
C = 0.8.
However, value of C smaller than 1 seem to be in contradiction with the Reggeon
unitarity condition~\cite{GM}. 

The differential cross section of elastic $\bar{p}p$ scattering at  
$\sqrt{s}$ = 62 GeV, $\sqrt{s}$ = 546 GeV, and $\sqrt{s}$ = 1.8 TeV calculated 
with the values of the parameters given in Eq.~(12) are presented in Fig.~4.
At the energy $\sqrt{s}$ = 62 GeV the theoretical curves are slightly below the
experimental points, probably due to the contribution of the $f$-Reggeon exchange,
that has not been accounted for in our calculations. 
\begin{figure}[htb]
\centering
\includegraphics[width=.75 \hsize]{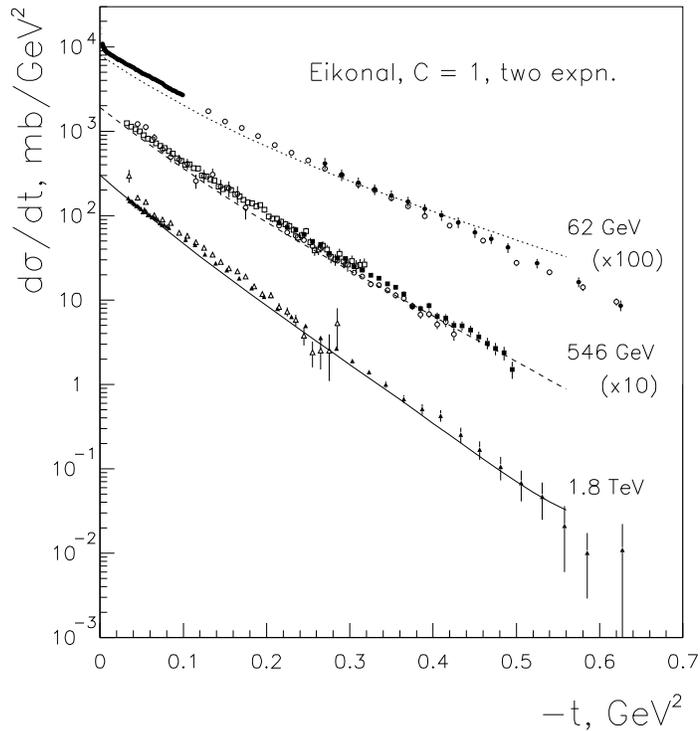}
\vskip -.6cm
\caption{\footnotesize
The differential cross section of elastic $\bar{p}p$ scattering at 
$\sqrt{s}$ = 62 GeV, \cite{Kwa,Am,Bak} $\sqrt{s}$ = 546 GeV \cite{Arn,Bat,Aug}, 
and $\sqrt{s}$ = 1.8 TeV \cite{Amo,Abe} calculated in the eikonal (C = 1) 
approach with the two-exponential parametrization of the function $\gamma(q^2)$ 
given in Eq.~(10).}
\end{figure}

The calculated values of total cross sections $\sigma^{tot}$,
of $d\sigma/dt(t=0)$, and of the slope of the elastic scattering cone 
parameter $B_{el}$ ($d\sigma/dt \sim exp(-B_{el}\cdot q^2)$) are presented 
in Table~1, together with the experimental data. It is necessary to note 
that the slope parameter was calculated in the interval $q^2 = 0-0.1$ GeV$^2$. 

\begin{center}
\begin{tabular}{|c||r|r|r|} \hline
$\sqrt{s}$  & $\sigma^{tot}$ (mb) & $d\sigma/dt(t=0)$ (mb/GeV$^2$) 
& $B_{el}$ (GeV$^{-2}$) \\ \hline
546 GeV & 60.6 \hspace{0.5cm} & 191\hspace{1.9cm} & 16.7\hspace{1.cm} \\
\cite{Boz} & $61.9 \pm 1.5$ & - \hspace{2.1cm} & - \hspace{1.cm} \\ 
\cite{Abe} & $61.3 \pm 0.9$ & $196 \pm 6$ \hspace{1.4cm}
& $15.35 \pm 0,19$ \\ \hline
1.8 TeV   & 76.2 \hspace{0.5cm} & 301\hspace{1.7cm} & 18.6\hspace{1.cm} \\
\cite{Abe} & $80.0 \pm 2.2$ & $335 \pm 19$  \hspace{1.2cm} 
& $16.98 \pm 0,25$ \\ 
\cite{Avi} & $71.7 \pm 2.0$ &  - \hspace{1.6cm} & - \hspace{1.6cm} \\ 
\hline
7 TeV & 97.6 \hspace{0.5cm} & 493 \hspace{1.57cm} & 21.2\hspace{1.cm} \\
\cite{TOT1a} & $98.3 \pm 2.8$ & - \hspace{1.8cm} & $20.1 \pm 0,4$  \\ \hline
\hline
\end{tabular}
\end{center}
\vskip-0.2cm
\noindent
Table 1. The comparison of the calculated values of total cross sections 
$\sigma^{tot}$, of $d\sigma/dt(t=0)$, and of the slope parameter $B$ with the
corresponding experimental data~\cite{TOT1a,Boz,Abe,Avi}.
\vskip0.2cm

The general energy dependence of the differential elastic $pp$ ($\bar{p}p$) 
cross sections is shown in Fig.~5. At the energy $\sqrt{s} = 62$~GeV some
contribution of $f$-Reggeon should be present.
\begin{figure}[htb]
\centering
\vskip -.7cm
\includegraphics[width=.65  \hsize]{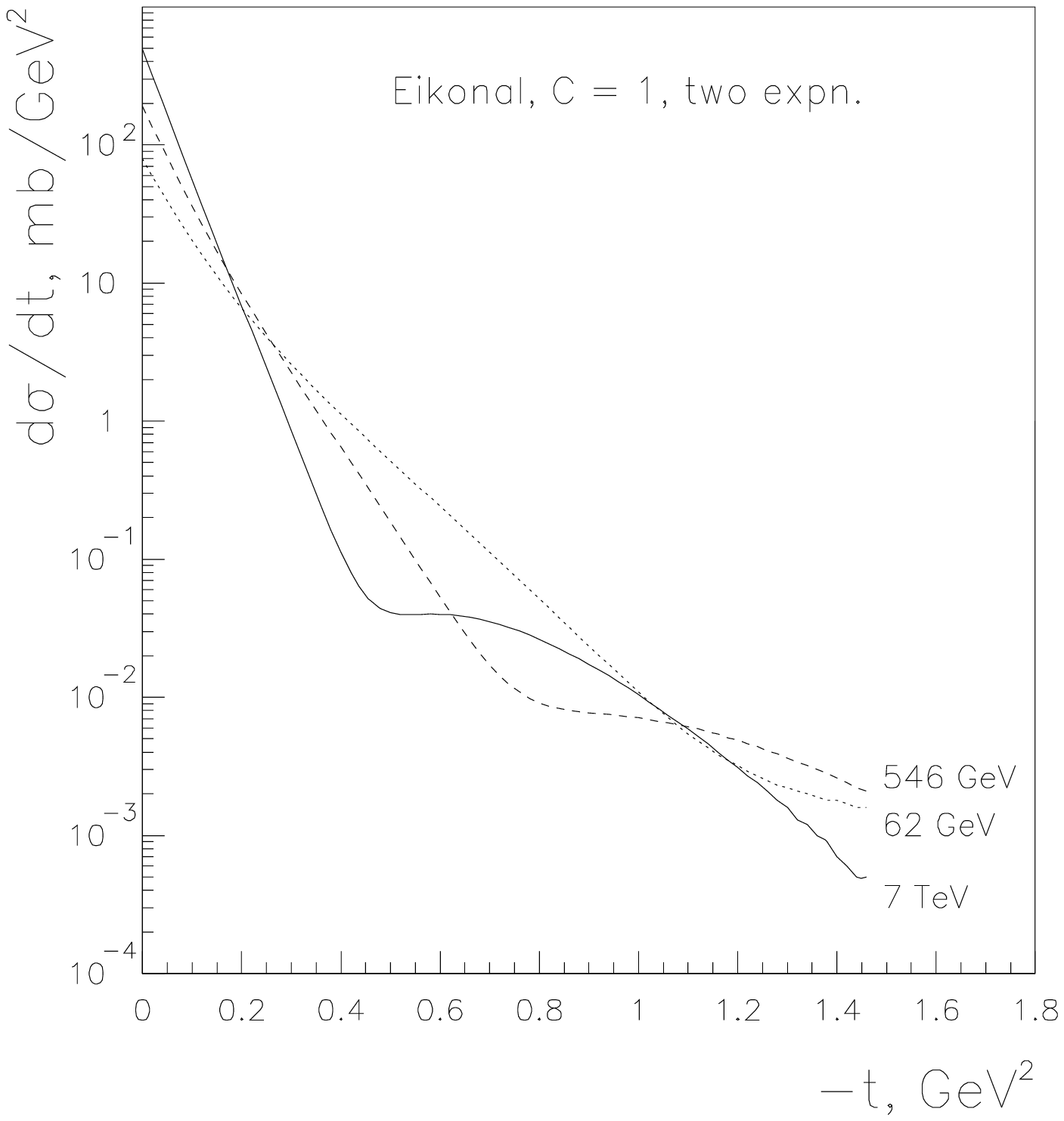}
\vskip -.6cm
\caption{\footnotesize
The differential cross section of elastic $pp$ scattering at  $\sqrt{s}$ = 
7 TeV (solid curve), $\sqrt{s}$ = 546 GeV (dottedd curve), and $\sqrt{s}$ = 
62 GeV (dashed curve) calculated in the eikonal approach (C = 1) with 
the two-exponential parametrization of the funtion $\gamma(q^2)$ given in
Eq.~(10).}
\end{figure}

However, in the complete Reggeon diagram technique~\cite{Gri} not only
Regge-poles and cuts, but also more complicated diagrams, e.g. the  so-called
enhanced diagrams, should be taken into account. In the numerical 
calculation of such diagrams some new uncertainties appear,
since the vertices of the coupling of multireggeon systems
are unknown. The common feature of such calculations results in the
additional increase of the Pomeron intercept $\alpha_P(0) = 1+\Delta$.

\section{Conclusion}

We obtain a general description of elastic $pp$ scattering that seems to be
successful, as one can see from Figs.~3 and 4, and from Table 1.
To do so we only use the three parameters shown in Eq.~(12), namely $\gamma$,
which 
determines the normalization of total $pp$ cross section, $\Delta$, which 
determines the increase of the total $pp$ cross section with energy, and
$\alpha'_P$ which determines the increase of the diffractive slope cone 
parameter. These parameters are practically not correlated. Another
three parameters, $a$, $R_1^2$, and $R_2^2$ are related to the geometrical 
shape of the proton and they should be determined from the experiment in the 
same way as we determine the geometrical shape of atomic nuclei. 

The exact values of the position of the diffractive dip and of the elastic
cross 
section in the dip-region strongly depend of the particular form, like those 
in Eqs.~(8) and (10), choosen to parmetrize the $q^2$-dependence of the 
Pomeron-nucleon coupling. With our oversimplified parametrization we did not 
succeed in describing the dip-region. On the other hand, it is sure this can 
be done by using a more complicated vertex $\gamma(q^2)$ with a larger number 
of parameters, as the parametrization used in~\cite{UG}, and that, strangely 
enough, shows a minimum at $b_t=0$, or the one in Eq.~(11c) of 
reference~\cite{DL}, which needs a not well justified additional term in
order to describe the dip.

{\bf Acknowledgements}

We are grateful to M.G. Ryskin for useful
discussions and comments. This paper was supported by Ministerio de 
Educaci\'on y Ciencia of Spain under the Spanish Consolider-Ingenio 2010
Programme CPAN (CSD2007-00042) and by project FPA 2005--01963, by Xunta de 
Galicia (Spain) and by University of Santiago de Compostela (Spain), and, in part,
by grant RFBR 11-02-00120-a.



\begin{thebibliography}{**}

\bibitem{TOT1a} G. Antchev et al., TOTEM Collaboration, Europhys. Lett. 
{\bf 96}, 21002 (2011).

\bibitem{DL} A. Donnachie and P.V. Landshoff, arXiv:1112.2485.

\bibitem{UG} V. Uzhinsky and A. Galoyan, arXiv:1111.4984.

\bibitem{RMK} M.G. Ryskin, A.D. Martin, and V.A. Khoze, arXiv:1201.6298.

\bibitem{Sel} O.V. Selyugin, arXiv:1201.4458.

\bibitem{Gri} V.N. Gribov,  ZhETF {\bf 53}, 657 (1967).

\bibitem{Volk} P.E. Volkovitsky, A.M. Lapidus, V.I. Lisin, and K.A.
Ter-Martirosyan, Yad. Fiz. {\bf 24}, 1237 (1976).

\bibitem{Kop} B.Z. Kopeliovich and  L.I. Lapidus, Sov. Phys JETP {\bf 44},
31 (1976).

\bibitem{Froi} M. Froissart, Phys. Rev. {\bf 123} (1961) 1053.

\bibitem{Kar1} K.A. Ter-Martirosyan, Yad. Fiz. {\bf 10}, 1047 (1969).

\bibitem{Kar3} K.A. Ter-Martirosyan, Phys. Lett. {\bf B44}, 377 (1973).


\bibitem{TOT2} G. Antchev et al., TOTEM Collaboration, Europhys. Lett. 
{\bf 95}, 1001 (2911) and arXiv: 1110.1385 [hep-ex]. 

\bibitem{GM} V.N. Gribov and A.A. Migdal, Yad. Fiz. {\bf 8}, 1002 (1968).

\bibitem{Kwa} N. Kwak et al., Phys. Lett. {\bf B58}, 233 (1975).

\bibitem{Am} U. Amaldi et al., Phys. Lett. {\bf B66}, 390 (1977).

\bibitem{Bak} A. Baksay et al., Nucl. Phys. {\bf B141}, 1 (1978).

\bibitem{Arn} G. Arnison et al., Phys. Lett. {\bf B128}, 336 (1982).

\bibitem{Bat} R. Battiston et al., Phys. Lett. {\bf B147}, 385 (1984).

\bibitem{Aug} C. Augier et al., Phys. Lett. {\bf B316}, 448 (1993).

\bibitem{Amo} N.A. Amos et al., Phys. Lett. {\bf B247}, 127 (1990).

\bibitem{Boz} M. Bozzo et al., Phys. Lett. {\bf B147}, 392 (1984).

\bibitem{Abe} F. Abe et al., Phys. Rev. {\bf D50}, 5518, 5550 (1994).

\bibitem{Avi} C. Avila et al., Phys. Lett. {\bf B445}, 419 (1999).


\end{thebibliography}
\end{document}